\RequirePackage{lineno}
\documentclass[aps,prc,twocolumn, floatfix]{revtex4}
\usepackage{graphicx}
\usepackage{dcolumn}
\usepackage{bm}

\begin{document}


\title{Probing the incompressibility of dense hadronic matter near QCD phase transition \\ in relativistic heavy-ion collisions}

\author{Zhi-Min Wu$^{1,2}$}
\author{Gao-Chan Yong$^{1,2}$}
\email[]{yonggaochan@impcas.ac.cn}

\affiliation{
$^1$Institute of Modern Physics, Chinese Academy of Sciences, Lanzhou 730000, China\\
$^2$School of Nuclear Science and Technology, University of Chinese Academy of Sciences, Beijing 100049, China
}

\begin{abstract}

Based on the extended hadronic transport model of relativistic heavy-ion collisions, the incompressibility of dense hadronic matter created in relativistic Au+Au heavy-ion collisions at $\sqrt{s_{NN}} = 3$ GeV is studied. By comparing experimental proton directed flow, productions of strange hadrons $\phi$, $K^{-}$ as well as their ratio $\phi/K^{-}$, proton high-order cumulants to the model calculations, a large incompressibility of dense hadronic matter is obtained from nucleon observabels while a rather small incompressibility is needed to fit the data of strange hadrons. This may indicate hadronic matter possesses different incompressibilities in different density regions, i.e., the incompressibility may become stiffer from saturation density to a certain baryon density and then turn to soft before reaching hadron-quark phase transition. The study also shows that the incompressibility significantly affects the critical baryon density of hadron-quark phase transition.

\end{abstract}

\maketitle

\section{Introduction}

One of the main goals of relativistic heavy-ion collisions is to understand the properties of nuclear matter under conditions of extreme energy and baryon density or to explore the phase diagram of hot and dense nuclear matters \cite{pr1,pr2,pr3}. A lot of circumstantial signals in relativistic heavy-ion collisions have suggested the formation of quark-gluon plasma (QGP) \cite{qgp1,qgp2}. Currently it is commonly considered that at small baryon chemical potential and high temperature, the transition from hadronic phase to QGP phase is a smooth crossover, whereas a first-order phase transition is expected at high baryon chemical potential region. Due to its fundamental importance, the exploration of quantum chromodynamics (QCD) phase diagram of strongly interacting nuclear matter is the current focus of many research activities worldwide, both theoretically and experimentally \cite{pt1,pt2,ptko22,guo2021,nara18}. Indeed, mapping the QCD phase diagram is the major scientific goal of the beam energy scan (BES) program in heavy-ion collisions \cite{pr2,bes19,besa,besb}. In astronomy, whether there is hadron-quark phase transition in neutron-stars (NSs) with central densities of several times nuclear saturation density is also of great interest in the study of neutron star structure \cite{akm1998,nature2020,liapj2020,hu2021,kojo2021,ran2021,xie2021} and gravitational-wave (GW) emission \cite{gw2019,gw2018, gw20182}. The study of the phase transition of QCD matter from earth to heaven is thought to have crucial implications toward an unprecedented understanding of the early and present universe \cite{ann2006}. It's worth noting that the exploration of dense matter theory for heavy-ion collisions and neutron stars has been one part of the new US Long-Range Plan in Nuclear Physics \cite{usplan2023}.

By constructing equation of state (EoS) of nuclear matter created in heavy-ion collisions and simulating relevant observables and comparing with experimental data, one can get related information on the stiffness of EoS as well as the hadron-quark phase transition of nuclear matter \cite{lipc2022,stj2022,dmy2022}. Previous studies on nucleon directed and elliptic flows by Danielewicz \emph{et al}. showed the possible change in compressibility with increasing baryon density \cite{dan2002}. Their studies seem to very different from the recent results by Nara \emph{et al}., the latter explained the colliding energy dependence of the proton directed flow with a single hadronic EoS \cite{nara22}. Fortunately, the recent STAR experiments clarified some facts on the properties of dense QCD matter \cite{v1flow2021,qgpa21}. Theoretically, the compressed baryon density reached in heavy-ion collisions depends on the EoS used and the effects of the EoS on relevant observables in heavy-ion collisions also depend on the compressed baryon density reached. Interplay of the compressed baryon density reached in heavy-ion collisions and the effects of the EoS without or with the hadron-quark phase transition at various densities complicates the question.

To probe the boundary of hadron-quark phase transition of nuclear matter, constraints could be obtained from the recent RHIC Beam Energy Scan-II program \cite{v1flow2021, comu2021,kpsi2021}. It is shown from different aspects that hadronic interactions dominate in matter created in relativistic Au+Au collisions at $\sqrt{s_{NN}}$ = 3 GeV. While it is not straightforward to obtain the baryon density at which there is no occurrence of hadron-quark phase transition in the Au+Au collisions at $\sqrt{s_{NN}}$ = 3 GeV. Our studies show that, to constrain the boundary of hadron-quark phase transition of nuclear matter, one first needs to constrain the incompressibility of dense hadronic matter near hadron-quark phase transition.

\section{The AMPT model with different modes}
A multi-phase transport (AMPT) model \cite{AMPT2005} is recently extended so that it can perform not only multi-phase transport simulations with both parton and hadron degrees of freedom but also pure hadron cascade with hadronic mean-field potentials \cite{cas2021}. As a Monte Carlo parton and hadron transport model, the AMPT model consists of four components, i.e., a fluctuating initial condition, partonic interactions, conversion from the partonic to the hadronic matter, and hadronic interactions \cite{AMPT2005}. The model has been extensively applied to heavy-ion collisions at RHIC and LHC energies \cite{nst2021}. In the AMPT model, $\pi$, $\rho$, $\omega$, $\eta$, $K$, $K^*$, $\phi$, $N$, $\Delta$, $N^*(1440)$, $N^*(1535)$, $\Lambda$, $\Sigma$, $\Xi$ and $\Omega$ are included \cite{deu2009}. In the pure hadron cascade model (AMPT-HC) \cite{cas2021}, the Woods-Saxon nucleon density distribution and local Thomas-Fermi approximation are used to initialize the position and momentum of each nucleon in colliding projectile and target. The parton degree of freedom is switched off. In addition to the usual elastic and inelastic collisions, hadron potentials with the test-particle method are applied to nucleons, baryon resonances, strangenesses as well as their antiparticles \cite{cas2021,yongrcas2022}. Since the form of the single nucleon potential at high momenta and high densities is still less known, to make minimum assumptions, here we use the density-dependent single nucleon mean-field potential
$U(\rho)=\alpha\frac{\rho}{\rho_0}+\beta(\frac{\rho}{\rho_0})^\gamma$ with $\alpha=(-29.81-46.9\frac{\kappa+44.73}{\kappa-166.32}){~\rm MeV}$,
$\beta=23.45\frac{\kappa+255.78}{\kappa-166.32}{~\rm MeV}$,
$\gamma=\frac{\kappa+44.73}{211.05}$ ($\rho_{0}$ and $\kappa$ stand for the saturation density and incompressibility of nuclear matter, respectively) to model the soft and stiff EoSs \cite{guo2021}.
As comparisons, the AMPT-SM mode with quark transport is also used in the present studies.

\section{Results and Discussions}
Recent observation of particle directed and elliptic flows in 10-40\% centrality for Au+Au collisions at $\sqrt{s_{NN}}$ = 3 GeV at RHIC is clear evidence that predominantly hadronic matter is created in such collisions and the QCD critical region, if created in heavy-ion collisions, could only exist at energies higher than 3 GeV \cite{v1flow2021, comu2021,kpsi2021}, for example, the recent STAR experimental studies on pion and proton elliptic flows show behavior which hints at constituent quark scaling (i.e., occurrence of quark matter) at $\sqrt{s_{NN}}$ = 4.5 GeV \cite{qgpa21}. The stiffness and the maximum compressed baryon density of the hadronic matter created in terrestrial laboratory have significant implications on the studies of the structure of neutron stars and the QCD phase diagram.

\begin{figure}[tbh]
\centering
\vspace{0.0cm}
\includegraphics[width=0.48\textwidth]{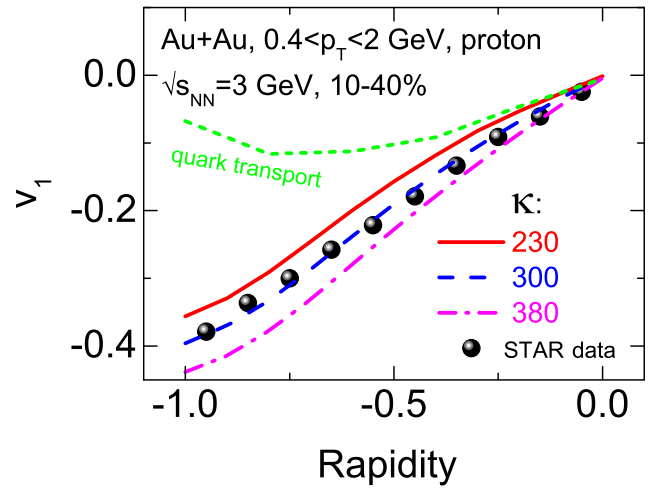}
\caption{Rapidity dependences of proton directed flow $v_{1}$ in 10-40\% centrality for Au+Au collisions at $\sqrt{s_{NN}}$ = 3 GeV given by the AMPT-HC mode with different EoSs and the quark transport AMPT-SM mode. The STAR data is taken from Ref.~\cite{v1flow2021}.} \label{v1flow}
\end{figure}
The slope of nucleon directed flow in semi-central heavy-ion collisions reflects the stiffness of nuclear matter created by the collisions of participant nucleons in target and projectile, thus is frequently used to probe the EoS or the QCD phase transition \cite{guo2021}.
Figure~\ref{v1flow} shows the proton directed flow $v_{1}$ as a function of rapidity given by the hadronic transport model AMPT-HC with different incompressibilities in the mean-field option. By comparing theoretical calculations with the variety of incompressibility to the experimental data \cite{v1flow2021}, one sees that a larger value of the incompressibility is needed to fit the data. The proton directed flow $v_{1}$ thus indicates a large incompressibility of dense hadronic matter created in Au+Au collisions at $\sqrt{s_{NN}}$ = 3 GeV. More specifically, an incompressibility coefficient of $k$ $\sim$ 300 MeV seems to be a good fit for the stiffness of the created hadronic matter in Au+Au collisions at $\sqrt{s_{NN}}$ = 3 GeV at RHIC. As comparison, the result of quark transport on the nucleon directed flow studied here indicates predominantly hadronic matter is created in such collisions.

\begin{figure}[tbh]
\centering
\vspace{0.0cm}
\includegraphics[height=6.6cm, width=0.48\textwidth]{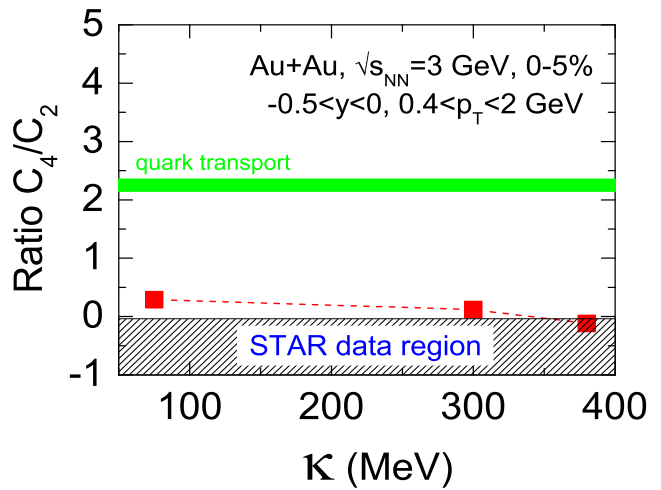}
\caption{The ratios of cumulants $C_{4}/C_{2}$ for proton in 0-5\% centrality for Au+Au collisions at $\sqrt{s_{NN}}$ = 3 GeV given by the AMPT-HC mode with different EoSs and the quark transport AMPT-SM mode. The hatched area stands for the STAR data, taken from Ref.~\cite{comu2021}.} \label{comu}
\end{figure}
Non-monotonic variation with $\sqrt{s_{NN}}$ of moments of the net-baryon number distribution, related to the correlation length and the susceptibilities of the system, is suggested as a signature for a critical point \cite{ks2}. It is interesting to see if such observable is also affected by the EoS of dense hadronic matter since the moments of the net-baryon number distribution reflect the properties of bulk nuclear system. The cumulant ratios of proton multiplicity distributions in Au+Au collisions at $\sqrt{s_{NN}}$ = 3 GeV were reported recently \cite{comu2021}. A suppression with respect to the Poisson baseline was observed in proton $C_{4}/C_{2}$ = -0.85 $\pm$ 0.09 $\pm$ 0.82 in the most central 0-5\% centrality collisions at 3 GeV. Since the quadratic variances is proportional
to approximate $\xi^{2}$, where $\xi$ is the correlation length, and the quartic one proportional
to $\xi^{7}$ \cite{c4c2}, the ratio of cumulants $C_{4}/C_{2} \sim \xi^{5}$. Because the soft EoS corresponds a larger correlation length $\xi$, the ratio of cumulants $C_{4}/C_{2}$ with the soft EoS should be larger than that with the stiff one. Figure~\ref{comu} shows the ratio of cumulants $C_{4}/C_{2}$ for proton in 0-5\% centrality for Au+Au collisions at $\sqrt{s_{NN}}$ = 3 GeV given by the AMPT-HC model with different incompressibilities in the mean-field option. It is seen that a soft EoS or a small incompressibility coefficient corresponds higher $C_{4}/C_{2}$ ratio for proton cumulants. To fit the data, one has to use a stiff EoS or a large incompressibility coefficient. More specifically, the incompressibility coefficient of the created hadronic matter in Au+Au collisions at $\sqrt{s_{NN}}$ = 3 GeV at RHIC seems to be $k > 300$ MeV. Again, the result of quark transport on the $C_{4}/C_{2}$ ratio studied here indicates predominantly hadronic matter is created in such collisions.

\begin{figure}[tbh]
\centering
\vspace{0.0cm}
\includegraphics[height=6.6cm, width=0.48\textwidth]{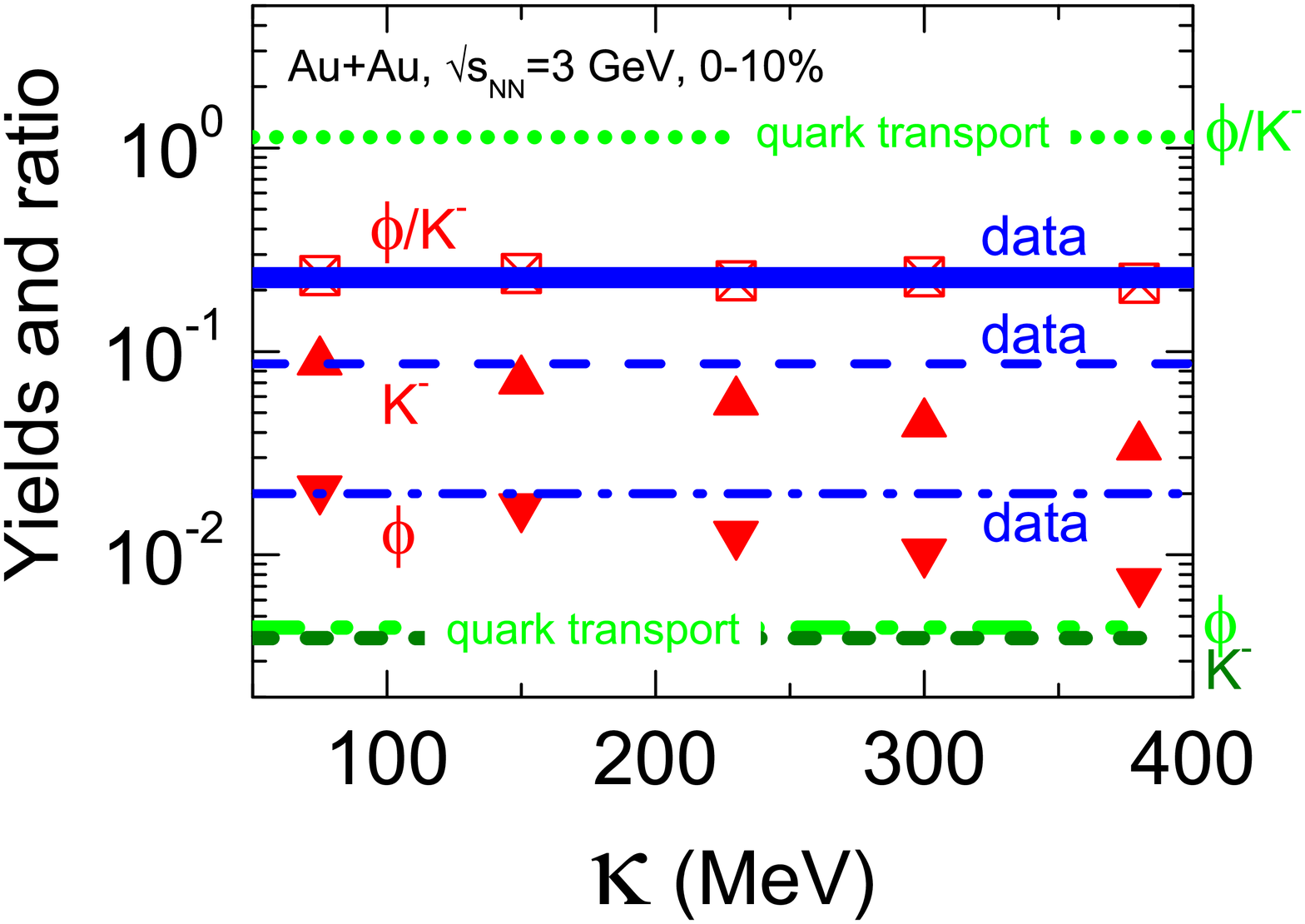}
\caption{Strange $\phi$ and $K^{-}$ productions and their ratios $\phi/K^{-}$ in 0-10\% centrality for Au+Au collisions at $\sqrt{s_{NN}}$ = 3 GeV given by the AMPT-HC mode with different EoSs and the quark transport AMPT-SM mode. The data is taken from Ref.~\cite{kpsi2021}.} \label{kpsi}
\end{figure}
Due to strangeness conservation, $K^{-}, \phi$ mesons, once produced, are rarely absorbed by the surrounding matter. The lack of final state interactions makes strange meson a penetrating probe for the EoS of dense matter produced in heavy-ion collisions. Also the strange meson is usually produced at maximum compression stage of nuclear collisions \cite{cas2021}. Figure~\ref{kpsi} shows the strange $\phi$ and $K^{-}$ productions and their ratio $\phi/K^{-}$ in 0-10\% centrality for Au+Au collisions at $\sqrt{s_{NN}}$ = 3 GeV given by the AMPT-HC model with different incompressibilities in the mean-field option. It is seen that the yields of the strange $\phi$ and $K^{-}$ productions are sensitive to the incompressibility in the mean-field option while their ratio $\phi/K^{-}$ is not affected by the variety of the incompressibility evidently. In heavy-ion collisions, the $K^{-}, \phi$ productions are mainly from baryon-baryon and meson-baryon or meson-meson collisions. The larger compression of the colliding nuclei, the more number of hadronic collisions occurs, thus more $K^{-}$ and $\phi$ are produced. Since the maximum compression reached is sensitive to the EoS, it is not surprising to see $K^{-}$ and $\phi$ productions are sensitive to the EoS. The soft EoS causes nuclear matter created in heavy-ion collisions to be more compressed, thus more mesons are produced. For the $K^{-}$'s, in the present study, we used an attractive $K^{-}$ potential in nuclear matter \cite{kaon1}. For the $\phi$ meson, although the in-medium corrections of the $\phi$ meson production were studied very recently \cite{phi1,phi2}, considering currently there is no specific $\phi$ single particle potential in the literature as kaon's, we did not use specific $\phi$ potential in the present study. Comparing to the experimental data \cite{kpsi2021}, it is found that a rather soft EoS or a fairly small incompressibility is needed to fit the data based on the model calculations. More specifically, the created hadronic matter in Au+Au collisions at $\sqrt{s_{NN}}$ = 3 GeV at RHIC seems to be soft, with the incompressibility coefficient of $75 < k < 150$ MeV. While the results of quark transport on the $\phi$ and $K^{-}$ productions and their ratio $\phi/K^{-}$ studied here evidently deviate the data, thus suggests predominantly hadronic matter is created in such collisions.

\begin{figure}[tbh]
\centering
\vspace{0.0cm}
\includegraphics[width=0.45\textwidth]{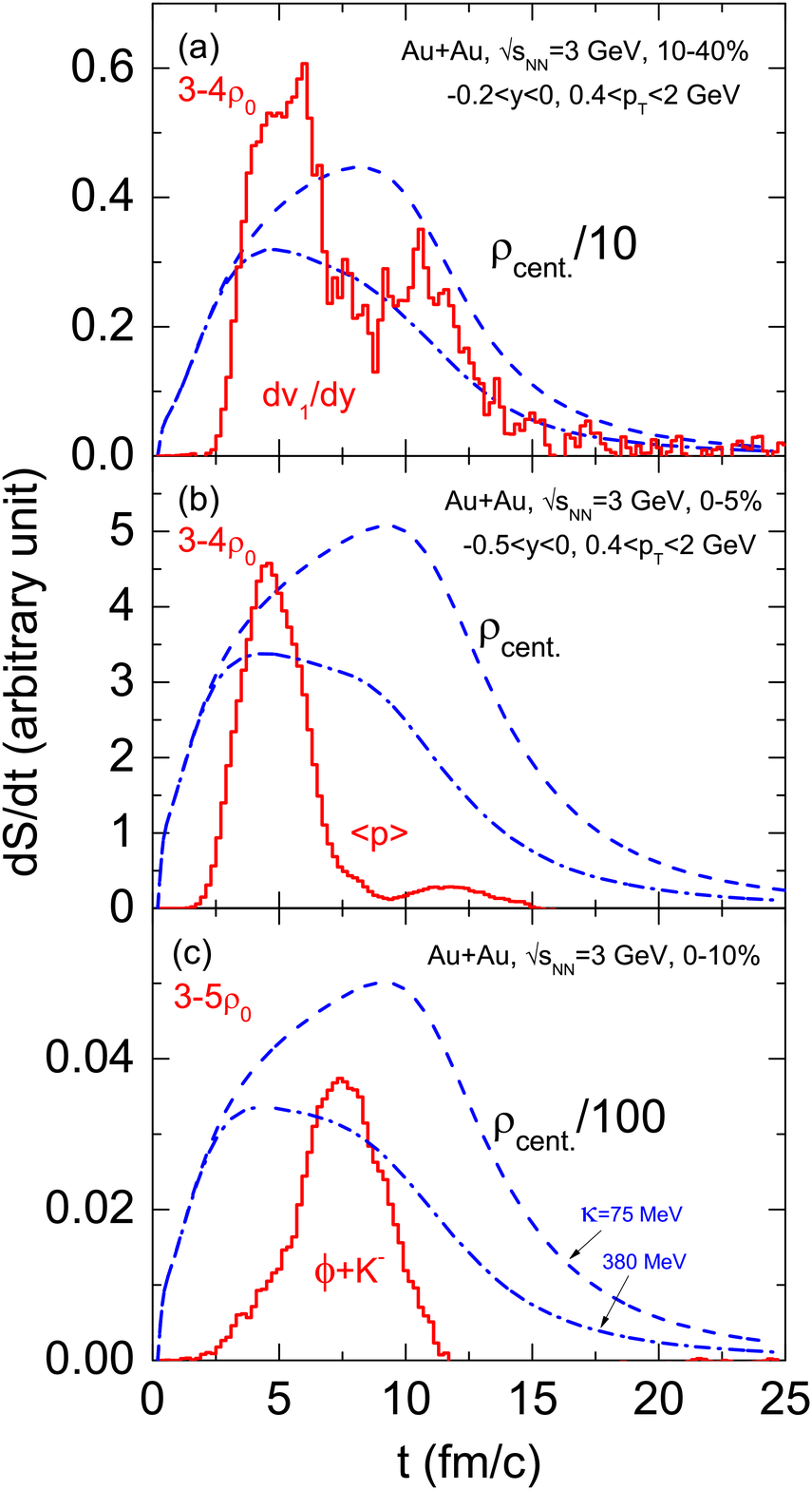}
\caption{Evolutions of the increment of sensitivity of incompressibility $dS/dt$ for different observables in Au+Au collisions at $\sqrt{s_{NN}}$ = 3 GeV simulated by the AMPT-HC mode. The maximum compressed baryon densities in a single central cell of 1 cubic Fermi with two different EoSs are also shown.} \label{dens}
\end{figure}
The obtained incompressibility or EoS of the hadronic matter created in Au+Au collisions at $\sqrt{s_{NN}}$ = 3 GeV varies with different observables as shown in Figures~1-3. This indicates complexity of the question of probing the properties of dense hadronic matter by using different observables in heavy-ion collisions.

To show why various EoSs or incompressibilities are obtained while using different observables via comparisons of theoretical results and experimental data in Au+Au collisions at $\sqrt{s_{NN}}$ = 3 GeV, specific density regions probed by different observables are demonstrated for Au+Au collisions at $\sqrt{s_{NN}}$ = 3 GeV. Figure~\ref{dens} shows sensitive density region of the incompressibility for different observables. Here the sensitivity $S = y(t)_{EoS1}-y(t)_{EoS2}$ with $y(t)_{EoS1}$ and $y(t)_{EoS2}$ being an observable's values with two different EoSs. The sensitivity increment is thus $dS/dt$, which reflecting the situation where and when the EoS plays a major role. It is seen that the sensitive density region of the incompressibility for both the proton directed flow $v_{1}$ and the proton multiplicity lies in the 3-4$\rho_{0}$ density region before maximum compression. Proton emission is mainly decided by the strength of the gradient force of nuclear mean-field potential which is proportional to the gradient of baryon density. It is thus not surprising to see the sensitivity of both the proton directed flow $v_{1}$ and the proton multiplicity to the incompressibility reaches maximum before the maximum compression of nuclear collisions.

While the strange $K^{-}, \phi$ are both secondary particles that are produced in the process of nucleons' and (or) mesons' multi-scatterings. So most strange mesons are produced around maximum compression of colliding nuclei. Although the larger gradient force of nuclear mean-field (or the larger gradient of baryon density) may speed up nucleon motion thus produce more mesons, these larger density gradients have minor effects compared with the maximum density compression. Therefore it is understandable that the sensitive density region of the incompressibility for strange mesons is around 3-5$\rho_{0}$ but still somewhat before the maximum nuclear compression as shown in the bottom panel of Figure~\ref{dens}. Since proton and strange meson observables are sensitive to various density regions of hadronic matter formed in heavy-ion collisions, it is not surprising to see different EoSs or incompressibilities are obtained from decoding relevant experimental data. In general, the present studies demonstrate that hadronic matter gradually softens as the density increases. This is exactly what we expected, since a hadron-quark phase transition would definitely occur at a certain baryon density point.

Another crucial implication of the present studies is that the critical density of hadron-quark phase transition of nuclear matter depends not only on the beam energy in heavy-ion collisions but on the incompressibility of dense matter. This is clearly shown in Figure~\ref{dens}. Supposing hadron-quark phase transition occurs in Au+Au collisions at $\sqrt{s_{NN}}$ = 3 GeV, the critical density of hadron-quark phase transition may lie in 3-5 times saturation density which depends on the incompressibility of dense matter near the phase transition.

\section{Conclusions}
In summary, by comparing the experimental data of nucleon and strange meson observables,  which have been carried out at RHIC, to the theoretical calculations of hadronic transport model in Au+Au collisions at $\sqrt{s_{NN}}$ = 3 GeV, various incompressibilities are obtained for hadronic matter created in heavy-ion collisions. The studies imply that hadronic matter gradually softens as the density increases from 3 to 5 times saturation density. Also the critical density of hadron-quark phase transition depends on the incompressibility of dense hadronic matter.

%
This work is supported by the National Natural Science Foundation of China under Grant No. 12275322 and
the Strategic Priority Research Program of Chinese Academy of Sciences with Grant No. XDB34030000.


\begin{thebibliography}{100}
%
\bibitem{pr1}Horst St\"{o}cker, Walter Greiner, Phys. Rep. {\bf 137}, 277 (1986).
\bibitem{pr2}M. M. Aggarwal \emph{et al}. (STAR Collaboration), arXiv: 1007.2613
\bibitem{pr3}Kenji Fukushima, Tetsuo Hatsuda, Rept. Prog. Phys. {\bf 74}, 014001 (2011).
\bibitem{qgp1}K. Adcox \emph{et al}. (PHENIX Collaboration), Nucl. Phys. A 757, 184 (2005).
\bibitem{qgp2}J. Adams \emph{et al}. (STAR Collaboration), Nucl. Phys. A 757, 102 (2005).
\bibitem{pt1}Adam Bzdak, ShinIchi Esumi, Volker Koch, Jinfeng Liao, Mikhail Stephanov, Nu Xu, Phys. Rep. {\bf 853}, 1 (2020).
\bibitem{pt2}X. Luo, N. Xu, Nucl. Sci. Tech. {\bf 28}, 112 (2017).
\bibitem{ptko22}Kai-Jia Sun, Wen-Hao Zhou, Lie-Wen Chen, Che Ming Ko, Feng Li, Rui Wang, Jun Xu, arXiv:2205.11010 (2022).
\bibitem{guo2021}Y. F. Guo, G. C. Yong, Phys. Lett. B {\bf 815}, 136138 (2021).
\bibitem{nara18}Y. Nara, H. Niemi, A. Ohnishi \emph{et al}., Eur. Phys. J. A {\bf 54}, 18 (2018).
\bibitem{bes19}G. Odyniec (for the STAR Collaboration), PoS {\bf CORFU2018}, 151 (2019).
\bibitem{besa}M. A. Stephanov, Prog. Theor. Phys. Suppl. {\bf 153}, 139 (2004).
\bibitem{besb}B. Mohanty, Nucl. Phys. A {\bf 830}, 899C (2009).
\bibitem{akm1998}A. Akmal, V. R. Pandharipande, and D. G. Ravenhall, Phys. Rev. C {\bf 58}, 1804 (1998).
\bibitem{nature2020}E. Annala, T. Gorda, A. Kurkela, J. N\"{a}ttil\"{a}, Nat. Phys. {\bf 16}, 907 (2020).
\bibitem{liapj2020}Z. Miao, A. Li, Z. Zhu, and S. Han, Astrophys. J. {\bf 904}, 103 (2020).
\bibitem{hu2021}Min Ju, Xuhao Wu, Fan Ji, Jinniu Hu, and Hong Shen, Phys. Rev. C {\bf 103}, 025809 (2021).
\bibitem{kojo2021}Toru Kojo, Defu Hou, Jude Okafor, and Hajime Togashi, Phys. Rev. D {\bf 104}, 063036 (2021).
\bibitem{ran2021}J. Ranjbar and M. Ghazanfari Mojarrad, Phys. Rev. C {\bf 104}, 045807 (2021).
\bibitem{xie2021}Wen-Jie Xie, Bao-An Li, Phys. Rev. C {\bf 103}, 035802 (2021).
\bibitem{gw2019}Andreas Bauswein, Niels-Uwe F. Bastian, David B. Blaschke, Katerina Chatziioannou, James A. Clark, Tobias Fischer, and Micaela Oertel, Phys. Rev. Lett. {\bf 122}, 061102 (2019).
\bibitem{gw2018}Soumi De, Daniel Finstad, James M. Lattimer, Duncan A. Brown, Edo Berger, and Christopher M. Biwer, Phys. Rev. Lett. {\bf 121}, 091102 (2018).
\bibitem{gw20182}Elias R. Most, L. Jens Papenfort, Veronica Dexheimer, Matthias Hanauske, Stefan Schramm, Horst St\"{o}cker, and Luciano Rezzolla, Phys. Rev. Lett. {\bf 122}, 061101 (2019).
\bibitem{ann2006}D. Boyanovsky, H. J. de Vega, D. J. Schwarz, Annu. Rev. Nucl. Part. Sci. {\bf 56}, 44 (2006).
\bibitem{usplan2023}Alessandro Lovato \emph{et al}.,  arXiv:2211.02224 (2022).
\bibitem{lipc2022}Pengcheng Li, Jan Steinheimer, Tom Reichert, Apiwit Kittiratpattana, Marcus Bleicher, Qingfeng Li, arXiv:2209.01413 (2022).
\bibitem{stj2022}Jan Steinheimer, Anton Motornenko, Agnieszka Sorensen, Yasushi Nara, Volker Koch, Marcus Bleicher, arXiv:2208.12091 (2022).
\bibitem{dmy2022}Dmytro Oliinychenko, Agnieszka Sorensen, Volker Koch, Larry McLerran, arXiv:2208.11996 (2022).
\bibitem{dan2002}P. Danielewicz, R. Lacey, and W. G. Lynch, Science {\bf 298}, 1592 (2002).
\bibitem{nara22}Yasushi Nara, Asanosuke Jinno, Koichi Murase, and Akira Ohnishi, Phys. Rev. C {\bf 106}, 044902 (2022).
\bibitem{v1flow2021}M. S. Abdallah \emph{et al}. (STAR Collaboration), Phys. Lett. B {\bf 827}, 137003 (2022).
\bibitem{qgpa21}M. S. Abdallah \emph{et al}. (STAR Collaboration), Phys. Rev. C {\bf 103}, 034908 (2021).
\bibitem{comu2021}M. S. Abdallah \emph{et al}. (STAR Collaboration), Phys. Rev. Lett. {\bf 128}, 202303 (2022).
\bibitem{kpsi2021}M. S. Abdallah \emph{et al}. (STAR Collaboration), Phys. Lett. B {\bf 831}, 137152 (2022).
%
\bibitem{AMPT2005}Zi-Wei Lin, Che Ming Ko, Bao-An Li, Bin Zhang, Subrata Pal, Phys. Rev. C {\bf 72}, 064901 (2005).
\bibitem{cas2021}Gao-Chan Yong, Zhi-Gang Xiao, Yuan Gao, Zi-Wei Lin, Phys. Lett. B {\bf 820}, 136521 (2021).
\bibitem{nst2021}Zi-Wei Lin, Liang Zheng, Nucl. Sci. Tech. {\bf 32}, 113 (2021).
\bibitem{deu2009}Yongseok Oh, Zi-Wei Lin, Che Ming Ko, Phys. Rev. C {\bf 80}, 064902 (2009).
\bibitem{yongrcas2022}Gao-Chan Yong, Bao-An Li, Zhi-Gang Xiao, and Zi-Wei Lin, Phys. Rev. C {\bf 106}, 024902 (2022).
\bibitem{ks2}J. Adam \emph{et al}. (STAR Collaboration), Phys. Rev. Lett. {\bf 126}, 092301 (2021).
\bibitem{c4c2}M. A. Stephanov, Phys. Rev. Lett. {\bf 102}, 032301 (2009).

\bibitem{kaon1}G. Q. Li, C.-H. Lee, G. E. Brown, Phys. Rev. Lett. {\bf 79}, 5214 (1997).
\bibitem{phi1}Taesoo Song, Joerg Aichelin, and Elena Bratkovskaya, Phys. Rev. C {\bf 106}, 024903 (2022).
\bibitem{phi2}A. Mishra, S. P. Misra, Eur. Phys. J. A {\bf 57}, 98 (2021).

\end{thebibliography}
\end{document}